# Generation of squeezed high-order harmonics

Matan Even Tzur[1], Michael Birk[2], Alexey Gorlach[2], Ido Kaminer[2], Michael Krüger[1], and Oren Cohen[1]

[1]Solid State Institute and Physics Department and Helen Diller Quantum Center, Technion-Israel Institute of Technology, Haifa 3200003, Israel.

[2]Solid State Institute and Department of Electrical and Computer Engineering and Helen Diller Quantum Center, Technion-Israel Institute of Technology, Haifa 3200003, Israel.

**Abstract**

For decades, most research of high harmonic generation (HHG) considered matter as quantum but light as classical, leaving the quantum-optical nature of the harmonics an open question. Here we explore the quantum properties of high harmonics. We derive a formula for the quantum state of the high harmonics, when driven by arbitrary quantum light states, and then explore specific cases of experimental relevance. Specifically, for a moderately squeezed pump, HHG driven by squeezed coherent light results in squeezed high harmonics. Harmonic squeezing is optimized by syncing ionization times with the pump's squeezing phase. Beyond this regime, as pump squeezing is increased, the harmonics initially acquire squeezed thermal photon statistics, and then occupy an intricate quantum state which strongly depends on the semi-classical nonlinear response function of the interacting system. Our results pave the way for generation of squeezed extreme-ultraviolet ultrashort pulses, and more generally, quantum frequency conversion into previously inaccessible spectral ranges, which may enable ultrasensitive attosecond metrology.

High harmonic generation [1,2] (HHG) occurs when intense light drives gases [1,2], liquids [3], solids [4], or plasma [5] to emit high-order harmonics of the driving field. Temporally, the HHG emission often consists of attosecond pulses [6]. In the gas phase, HHG is understood in terms of the three-step model [7–9]. An initially bound electron undergoes laser-induced tunnel ionization, then accelerates in the continuum under the influence of the oscillating laser field, and finally, it recombines with its parent ion, releasing its kinetic and potential energy as a high-energy photon. Notably, HHG played a critical role in the development of attosecond science [10], which temporally resolves structure and dynamics of matter with attosecond resolution by measuring optical emission and photoelectron spectra in various geometries [11–18]. Additionally, HHG is useful for a wide range of spectroscopic applications, e.g., for probing topological phase transitions [19], molecular chirality [20,21], ring currents [22] and symmetry breaking [23,24]. However, despite its wide range of scientific and technological applications, it remains unclear if HHG, like $\chi^{(2,3)}$ nonlinearities, can be used for frequency-converting quantum light without altering its quantum state, a process known as quantum frequency conversion [25,26] (QFC). While QFC based on low-order $\chi^{(2,3)}$ nonlinearities is successful, these nonlinearities generally do not support extreme up conversion of light. Consequently, transferring quantum correlations from the near-infrared (NIR) to the extreme ultraviolet (XUV) spectral range or from mid-infrared (MIR) to ultraviolet (UV) remains inaccessible. HHG represents the natural next step in this context. Recently, quantum optical aspects of both the outgoing pump and the emitted high harmonics have been extensively researched [27–31]. Most attention was invested in the special case of a driving field being initially a Glauber coherent state [32–40] – i.e., a classical drive. Indeed, to date, HHG experiments were only driven by coherent-state (i.e., "classical") light fields corresponding to the emission of a laser. This is because only coherent-state light was accessible with high intensities and ultrashort pulse durations. This situation is changing rapidly, as ultrashort pulses of intense non-classical light are now well-established drivers of nonlinear optics [41,42] and are starting to approach the regime of strong-field physics [41,43–48]. Indeed, recent work [49,50] suggested that HHG driven by squeezed vacuum pulses should already be within reach by combining the most efficient HHG geometries [51] with tabletop squeezed vacuum beamlines [52]. These works predicted an extended HHG cutoff generated by squeezed vacuum [49], and novel control over electron trajectories [50] through an effective photon statistics force. Nevertheless, the quantum state of the emitted high harmonics remains unknown. By uncovering the relationship between the quantum state of the pump, material parameters, and the HHG emission, it may be possible to harness HHG for QFC, extending it into currently inaccessible spectral ranges.

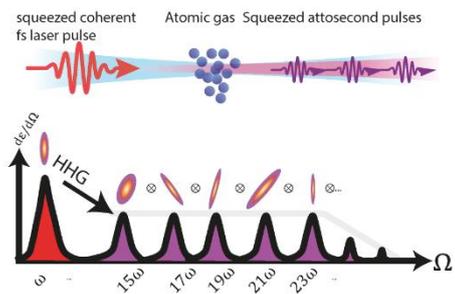

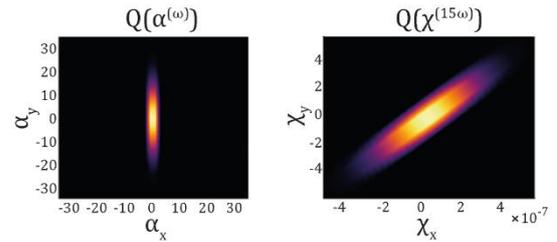

**Figure 1. Quantum state transfer between spectral ranges using high harmonic generation.** (a) Illustration of HHG driven by amplitude squeezed light (b) Husimi distribution of the pump at frequency $\omega$ (left) and the 15th harmonic at frequency $15\omega$ (right), calculated using Eq. (9) and the strong-field approximation.

Here we explore the quantum state of high harmonics and show that when driven by quantum light, the HHG emission inherits quantum features from the pump. Our work applies to any arbitrary quantum state of light, providing an analytical formula for the quantum state of the harmonics. After laying out the general formalism, we focus on two specific examples that are experimentally accessible. First, we consider HHG in atomic gas driven by squeezed coherent light, and show that the harmonics may exhibit squeezing when the pump is squeezed, i.e., HHG may support QFC of squeezed light. Importantly, absolute squeezing of high harmonics (i.e., beyond shot noise) is achieved only for a limited range of pump squeezing, because the contribution of the anti-squeezed quadrature to the nonlinear emission is mapped to both quadratures of the harmonics. Our analysis reveals how the quantum state of the emitted harmonics depends on the underlying electronic trajectories, showing that harmonic squeezing is optimized when their ionization time is synchronized with the squeezed phase of the pump. In the second example, we consider HHG driven by bright squeezed vacuum. We find that in this case, the harmonics occupy an intricate quantum state. We explain the key features of this state using an analogy to a winding trail down a mountain. Our results pave the way towards the generation of squeezed ultrashort XUV light by means of extremely nonlinear QFC, which may bring below-shot-noise metrology to the attosecond timescale and x-ray spectral range.

**Quasi-probability distributions of the high harmonic emission**

We begin by deriving the density matrix of the HHG emission due to interaction of an atom with light at frequency $\omega$, whose quantum state is specified through its Positive P representation $P_\omega(\alpha, \beta)$ [53]. We note that such a decomposition exists for any single mode quantum state, it is nonnegative, and it is normalized to 1. The density matrix is given by

$$\hat{\rho}_{light}(t=0) = \int d^2\alpha d^2\beta \cdot P_\omega(\alpha, \beta^*) \frac{|\alpha\rangle\langle\beta|}{\langle\beta|\alpha\rangle} \bigotimes_{q=2}^{q_{cutoff}} |0_{q\omega}\rangle\langle 0_{q\omega}| \qquad (1)$$

where $t = 0$ is before the start of the interaction. Here, $|\alpha\rangle$ and $|\beta\rangle$ are Glauber coherent states at frequency $\omega$, and $|0_{q\omega}\rangle$ are vacuum modes at frequencies $q\omega$ where $q$ is an integer. Assuming that the material system initially fully occupies the ground state of the atom $|g\rangle$, the initial condition for the joint light matter density matrix is given by $\hat{\rho}(0) = \hat{\rho}_A(0) \otimes \hat{\rho}_{light}(0)$ where $\hat{\rho}_A(0) = |g\rangle\langle g|$. The Schrodinger equation for $\hat{\rho}(t)$ is governed by the Hamiltonian $\hat{H} = \hat{H}_A - e\hat{\boldsymbol{d}} \cdot \hat{\boldsymbol{E}}(t)$, where $\hat{H}_A$ is the atomic Hamiltonian, $\hat{\boldsymbol{d}}$ is the dipole moment expectation value, and $\hat{\boldsymbol{E}}(t) = i\sum_{\boldsymbol{k},\sigma}\sqrt{\frac{\hbar\omega}{2V\varepsilon_0}}\boldsymbol{\varepsilon}(\hat{a}_{\boldsymbol{k}\sigma}e^{-i\omega t} - \hat{a}^\dagger_{\boldsymbol{k}\sigma}e^{i\omega t})$ is the electric field operator. Using the linearity of the Schrödinger equation, we decompose $\hat{\rho}(t)$ as

$$\hat{\rho}(t) \equiv \int d^2\alpha d^2\beta \cdot P_\omega(\alpha, \beta^*) \hat{\rho}_{\alpha\beta}(t) \qquad (2)$$

where $\hat{\rho}_{\alpha\beta}(t)$ is defined as the component of the density matrix whose initial condition is $\hat{\rho}_{\alpha\beta}(0) = \hat{\rho}_A(0) \otimes \frac{|\alpha\rangle\langle\beta|}{\langle\beta|\alpha\rangle} \otimes_{q=2}^{q_{cutoff}} |0_{q\omega}\rangle\langle 0_{q\omega}|$. Next, we use coherent shift operators to transform the initial condition of each $\hat{\rho}_{\alpha\beta}$ into parameters in its equation of motion. Using $\widehat{D}^\dagger(\alpha), \widehat{D}(\beta)$, we define $\tilde{\rho}_{\alpha\beta}(t) = \widehat{D}^\dagger(\alpha)\hat{\rho}_{\alpha\beta}(t)\widehat{D}(\beta)$ which satisfies the initial condition $\langle\beta|\alpha\rangle\tilde{\rho}_{\alpha\beta}(0) = \hat{\rho}_A(0) \otimes_{q=1}^{q_{cutoff}} |0_{q\omega}\rangle\langle 0_{q\omega}|$ where the tensor product now begins at $q = 1$ (in contrast to Eq. (1)). Using the Schrödinger equation, Eq. (2), and the equations above, the time evolution of $\tilde{\rho}_{\alpha\beta}$ is given by

$$i\hbar \frac{\partial \tilde{\rho}_{\alpha\beta}}{\partial t} = \widehat{H}_\alpha \tilde{\rho}_{\alpha\beta} - \tilde{\rho}_{\alpha\beta} \widehat{H}_\beta - [\widehat{\boldsymbol{d}} \cdot \widehat{\boldsymbol{E}}, \tilde{\rho}_{\alpha\beta}]$$

$$\widehat{H}_\alpha \equiv H_A - \widehat{\boldsymbol{d}} \cdot \boldsymbol{E}_\alpha(t) \tag{3}$$

$$\widehat{H}_\beta \equiv H_A - \widehat{\boldsymbol{d}} \cdot \boldsymbol{E}_\beta(t)$$

where $\boldsymbol{E}_\alpha(t) = \langle \alpha | \widehat{\boldsymbol{E}} | \alpha \rangle$, $\boldsymbol{E}_\beta(t) = \langle \beta | \widehat{\boldsymbol{E}} | \beta \rangle$, and we have taken $e = 1$. We introduce the ansatz $\langle \beta | \alpha \rangle \tilde{\rho}_{\alpha\beta}(t) = |\phi_\alpha(t)\rangle |\chi_\alpha(t)\rangle \otimes \langle \phi_\beta(t) | \langle \chi_\beta(t) |$ where $|\phi_{\alpha,\beta}\rangle$ are electronic wavefunctions, and $|\chi_{\alpha,\beta}\rangle$ are photonic wavefunctions which span multiple modes of the electromagnetic field. Using this ansatz, Eq. (3) is reformulated as

$$i\hbar \frac{\partial |\phi_\alpha\rangle |\chi_\alpha\rangle}{\partial t} = H_\alpha |\phi_\alpha\rangle |\chi_\alpha\rangle - \widehat{\boldsymbol{d}} \cdot \widehat{\boldsymbol{E}} |\phi_\alpha\rangle |\chi_\alpha\rangle \tag{4}$$

Identical equations apply for $|\phi_\beta\rangle |\chi_\beta\rangle$. Equation (4) gives physical meaning to the constituents of the ansatz $|\phi_{\alpha,\beta}\rangle$ and $|\chi_{\alpha,\beta}\rangle$: they are the electronic and photonic parts of the solution of the joint light-matter Schrodinger equation for HHG driven by a coherent state $|\alpha\rangle$ ($|\beta\rangle$) of light. In this equation, the term $H_\alpha |\phi_\alpha\rangle |\chi_\alpha\rangle$ expresses the classical driving field $E_\alpha(t)$ carried by the coherent state $|\alpha\rangle$, and $-\widehat{\boldsymbol{d}} \cdot \widehat{\boldsymbol{E}} |\phi_\alpha\rangle |\chi_\alpha\rangle$ expresses the coupling of the classically driven electron to the quantized modes of the electromagnetic field. Neglecting all action of the quantized modes on the electron, its time dependent state is given by the semi-classical Schrödinger equation $i\hbar \partial_t |\phi_\alpha\rangle = H_\alpha |\phi_\alpha\rangle$. Physically, this assumption means that photon-emission recoil is negligible, and that the interaction is much faster than any spontaneous emission time of the atomic system. Then, $|\chi_\alpha\rangle$ is given (up to a quantum phase factor which we neglect) by [54]:

$$|\chi_\alpha\rangle = |\alpha_\omega + \chi_\omega(\alpha)\rangle \bigotimes_{q=2}^{q_{cutoff}} |\chi_{q\omega}(\alpha)\rangle \tag{5}$$

in which $|\chi_q(\alpha)\rangle$ is a coherent state with displacement $\chi_q(\alpha) = -iN\sqrt{q}\epsilon^{(1)} \langle d_\alpha(q\omega)\rangle$ [54]. Here, $\epsilon^{(1)} = \sqrt{\hbar\omega/2\epsilon_0 V}$, $N$ is the number of phase-matched atoms, and $\langle d_\alpha(q\omega)\rangle = \int_{-\infty}^{\infty} dt \langle \phi_\alpha(t) | \widehat{d} | \phi_\alpha(t) \rangle e^{iq\omega t}$ is the semi-classical dipole moment expectation value of the electronic wavepacket driven by the $|\alpha\rangle$ coherent state $|\phi_\alpha(t)\rangle$. Hence, at time $t$, the density matrix of the system is given by

$$\hat{\rho}(t) = \int d^2\alpha d^2\beta \cdot P_\omega(\alpha, \beta^*) |\phi_\alpha(t)\rangle \langle \phi_\beta(t)| \otimes \frac{|\chi_\alpha\rangle\langle\chi_\beta|}{\langle\beta|\alpha\rangle} \tag{6}$$

The final step to obtain the density matrix of light is to trace out all electronic degrees of freedom. For this, we note that by definition $|\phi_\alpha\rangle |\chi_\alpha\rangle$ is the solution of the complete light-matter Schrödinger equation with initial condition $|g\rangle |\alpha\rangle$ (Eq.(4)). Thus, we have:

$$\langle\chi_\beta|\chi_\alpha\rangle Tr_E(|\phi_\alpha\rangle\langle\phi_\beta|) = Tr_{E,F}(|\phi_\alpha\rangle|\chi_\alpha\rangle\langle\chi_\beta|\langle\phi_\beta|)$$
$$= Tr_{E,F}\left(U_{AF}^\dagger(t,t')|g\rangle|\alpha\rangle\langle\beta|\langle g|U_{AF}(t,t')\right) = \langle\beta|\alpha\rangle \quad (7)$$

where $U_{AF}(t,t')$ is the unitary time evolution operator of the light matter system. Therefore, the density matrix of light after the interaction is given by

$$\rho_{light} = \int d^2\alpha d^2\beta \cdot P(\alpha,\beta^*)\frac{|\chi_\alpha\rangle\langle\chi_\beta|}{\langle\chi_\beta|\chi_\alpha\rangle} \quad (8)$$

Equation (8) is a central result of the paper. It gives the complete state of light in HHG driven by quantum light. It shows how the quantum state of the driver $P_\omega(\alpha,\beta^*)$ mixes with the semi-classical nonlinear response function of the system $\chi(\alpha)$ which may be calculated by solving the semi-classical time dependent Schrodinger equation [55] or by employing the strong-field approximation theory [8]. Notably, equation (8) leads to the Glauber-Sudarshan and Husimi distribution of harmonic $q\omega$ (see SI for more details of the derivation and a formula for the positive $P$ representation of harmonic $q\omega$):

$$P_{q\omega}(\chi_{q\omega}) = J(\chi_{q\omega}) \cdot P_\omega\left(\alpha(\chi_q)\right)$$
$$Q_{q\omega}(\chi_q) = J^2(\chi_{q\omega}) \cdot Q_\omega\left(\alpha(\chi_q)\right) \quad (9)$$

$$J(\chi_q) = \left|\frac{\partial(\alpha_x,\alpha_y)}{\partial(\chi_{qx},\chi_{qy})}\right| = \left|\partial_{\chi_x}\alpha_x\partial_{\chi_y}\alpha_y - \partial_{\chi_y}\alpha_x\partial_{\chi_x}\alpha_y\right|$$

Here, $\alpha(\chi_q)$ is the inverse function of $\chi_q(\alpha)$ and the $J(\chi_q)$ is the Jacobian of the transformation $\alpha \to \chi_q$. Equation (9) expresses explicitly the quantum state of the frequency component $q\omega$, and is therefore another central result of the paper. It shows that the quantum state of the driving field is mapped to the quantum state of emitted harmonic through the classical nonlinear response function $\chi_q(\alpha_\omega)$. For example, Fig. 1(b) shows the $x$ and $y$ quadratures of the 15th harmonic, calculated using the strong-field approximation, assuming an atomic system with a single structureless bound state.

Before proceeding to examine specific cases, we briefly revisit and sum up the approximations leading to Eq. (9). Firstly, in writing Eq. (3), we have implicitly assumed that the quasi-probability distribution of the pump is independent of time, which means, for example, that it is not depleted during the HHG interaction. Additionally, in writing Eq. (5), we have neglected a quantum phase prefactor $e^{-i\zeta(\alpha)}$ which in principle should be included in the state $|\chi_\alpha\rangle$. This assumption is justified as long as $\zeta(\alpha) - \zeta(\beta)$ is flat within the sampled phase-space area of $P_\omega(\alpha,\beta^*)$, i.e., that the term linear in $P_\omega(\alpha,\beta)$ does not contribute significantly to the emission if $|\alpha - \beta| \gg 1$. This assumption is reasonable in most cases as indicated by the relation between the positive P and Husimi representations of an arbitrary quantum state $P(\alpha,\beta^*) = \frac{1}{4\pi}\exp\left(-\frac{|\alpha-\beta|^2}{4}\right)Q\left(\frac{\alpha+\beta}{2}\right)$. Finally, we have assumed that HHG maps a coherent driver state to a coherent harmonic state. While this approximation is widely used in the literature [31,32,35], it is important to emphasize that it was not yet tested experimentally and may not always hold, for example, when the emitters are correlated [27].

## High harmonic generation driven by squeezed coherent light

In this section, we consider the experimentally feasible example of HHG driven by squeezed coherent light. To visualize the squeezed coherent pump, it is instructive to consider one possible way of its generation, by using a beam splitter to superimpose a bright coherent state with a dim squeezed vacuum field [56]. By varying the relative phase between the bright coherent state and the squeezed vacuum beams, one may continuously transition between phase-squeezed light (with a certain phase but uncertain amplitude) and amplitude squeezed light (with a certain amplitude yet uncertain phase). The squeezed coherent state of the pump is denoted by $|\gamma, r\rangle$ where $\gamma$ and $r$ are its dimensionless coherent state and squeezing parameters, respectively. The Husimi distribution of this state is given by [53]

$$Q_\omega^{(SC)}(\alpha) = \frac{1}{\pi \cosh(r)} \exp\left[-\frac{2(\alpha_y - \gamma_y)^2}{1 + e^{2r}} - \frac{2(\alpha_x - \gamma_x)^2}{1 + e^{-2r}}\right] \qquad (10)$$

Each value of $\alpha = \alpha_x + i\alpha_y$ in the distribution $Q_\omega^{(SC)}(\alpha)$ corresponds to a coherent state carrying a classical electromagnetic field $E_\alpha(t) = 2\epsilon^{(1)}(\alpha_x \sin(\omega t) + \alpha_y \cos(\omega t))$, where $\epsilon^{(1)} \approx 100$ V/cm in free space, is the amplitude of electric field vacuum fluctuations [57]. To calculate the quantum state of the emitted harmonics, we solve for $\chi^{(q)}(\alpha)$ using the strong-field approximation theory [8]. Then, using the obtained $\chi^{(q)}(\alpha)$ and Eqs. (9) and (10), we construct the Husimi distribution of each harmonic. Figure 1(b) shows the result of this construction for harmonic 15 driven by squeezed coherent light. While the harmonic is squeezed, it does not preserve the squeezing phase of the pump or the pump's degree of squeezing. Additionally, the degree and phase of squeezing exhibit dispersion with harmonic order (see Fig. 1 (a) for an illustration).

To explore this further, we derive an analytical formula for the quadrature variance $\Delta X_\theta^2$ for harmonic $q\omega$ generated by a squeezed coherent state $|\gamma, r\rangle$ using the Glauber-Sudarshan $P_\omega(\alpha)$ representation version of Eq. (9). We outline below the key steps and approximations of the derivation (see section III of the SI for detailed derivation). The quadrature $\hat{X}_\theta$ of the $q\omega$ harmonic is defined by $\hat{X}_\theta = \hat{X}\cos(\theta) + \hat{P}\sin(\theta)$ where $\hat{X} = \hat{a}_{q\omega} + \hat{a}^\dagger_{q\omega}$, $\hat{P} = -i(\hat{a}_{q\omega} - \hat{a}^\dagger_{q\omega})$, and $\hat{a}_{q\omega}$ is an annihilation operator for the $q\omega$ mode. For the $\hat{X}$ quadrature ($\theta = 0$), Eq.(9) shows $\langle \hat{X}^2 \rangle = 1 + 4\int P_\omega(\alpha)\chi_x^2(\alpha)\,d^2\alpha$ where $\chi_x = \Re\{\chi\}$. Using the strong-field approximation, neglecting long trajectories, and setting $I_p = 0$ (SI, section II), we approximate $\chi_x(\alpha) \approx \chi_{qx}(\gamma)\cos(\sigma_{q\omega}) - \chi_{qy}(\gamma)\sin(\sigma_{q\omega})$ where $\sigma_{q\omega} = A_{q\omega}(\alpha_x - \gamma_x) + B_{q\omega}(\alpha_y - \gamma_y)$ is a real valued parameter. Here, $\chi_{qx,y}(\gamma)$ are quadrature amplitudes of the $q\omega$ state generated by a coherent state $|\gamma\rangle$. The coefficients $A, B_{q\omega}$ are given in explicit form in the SI section II, and are functions of canonical momentum $p$, ionization time $t_0$, and recombination time $t$ of harmonic $q\omega$. Approximating $\sigma \ll 1$ in the phase space area covered by $P_\omega(\alpha)$, we obtain (to lowest order in $\epsilon^{(1)}$) $\Delta \hat{X}^2 = 1 + 4(\sqrt{N}\epsilon^{(1)})^4 q\langle d_{\gamma x}(q\omega)\rangle^2 \left(A_{q\omega}^2(e^{-2r} - 1) + B_{q\omega}^2(e^{2r} - 1)\right)$ in which $\langle d_{\gamma x}(q\omega)\rangle$ is the real part of the semi-classical dipole moment $\langle d_\gamma(q\omega)\rangle$. Through similar derivations of $\Delta \hat{P}^2$ and $\langle XP + PX \rangle - 2\langle X \rangle\langle P \rangle$, we arrive at the final expression for $\Delta X_\theta$ (to leading order in $\epsilon^{(1)}$):

$$\Delta \hat{X}_\theta^2 = 1 + 4\epsilon^{(1)4}N^2 q(\langle d_{\gamma x}(q\omega)\rangle\cos(\theta) + \langle d_{\gamma y}(q\omega)\rangle\sin(\theta))^2 \left(A_{q\omega}^2(e^{-2r} - 1) + B_{q\omega}^2(e^{2r} - 1)\right) \qquad (11)$$

in which $\theta$ is quadrature angle, $N$ is the number of interacting atoms, $q$ is the harmonic order, $\epsilon^{(1)}$ is the single photon amplitude at frequency $\omega$, $\langle d_{\gamma x,y}(q\omega)\rangle$ are the real and imaginary parts of the semi-classical dipole moment $\langle d_\gamma(q\omega)\rangle$ of the harmonic $q\omega$ generated by a coherent state $|\gamma\rangle$ (semi-classical), $A, B_{q\omega}$ are functions of $[p, t_0, t_1]_{q\omega}$ and $r$ is a dimensionless squeezing parameter. Equation (11) is another central result of the paper.

Figure 2(a,b) depicts the quadrature variances calculated with Eq. (11), plotted against the pump squeezing parameter $r$ for amplitude and phase squeezed pumps respectively, with intensity $2 \times 10^{14}\,\text{Watt/cm}^2$, and for short trajectories only. Notably, we observe that absolute squeezing (i.e., squeezing beyond the level of vacuum fluctuations) is limited to a finite range of $r$ (this can be seen by taking the limit $r \to \infty$ in Eq. (11)). Beyond this range, we find that the harmonics exhibit thermal statistics. From a physical standpoint, the "thermalization" of harmonics occurs because the anti-squeezing of the pump is mapped to both quadratures of the emitted harmonics, eventually spoiling absolute squeezing altogether. For low pump squeezing, Fig. 2(a) shows that for an amplitude-squeezed pump ($E_{\gamma_x} = 0.0756\,\text{a.u.}, E_{\gamma_y} = 0$), the lowest harmonics do not exhibit squeezing and their quadrature variances exceeds 1. Higher harmonics do exhibit absolute squeezing (quadrature variance $< 1$) which increases with harmonic order. To shed light on this pattern, it is instructive to consider that for an amplitude squeezed pump, the photon fluctuations are squeezed at the peak of the field, where the ionization time window is located, and gradually become anti-squeezed at later times. Additionally, we note that for short trajectories, the highest harmonics correspond to ionization times closest to the peak of the field, where the pump's fluctuations are minimal. Consequently, electrons that emit higher harmonics experience the least noise during their tunneling step, which eventually leads to squeezed high harmonics. This reasoning is also consistent with Fig. 2(b) which shows the harmonic quadrature variances for a phase squeezed pump. Here, field fluctuations are anti-squeezed within the ionization window and consequently none of the harmonics exhibit squeezing. The highest harmonics exhibit the highest quadrature variance, corresponding to their ionization time closest to the peak of the field, where the pump maximally fluctuates.

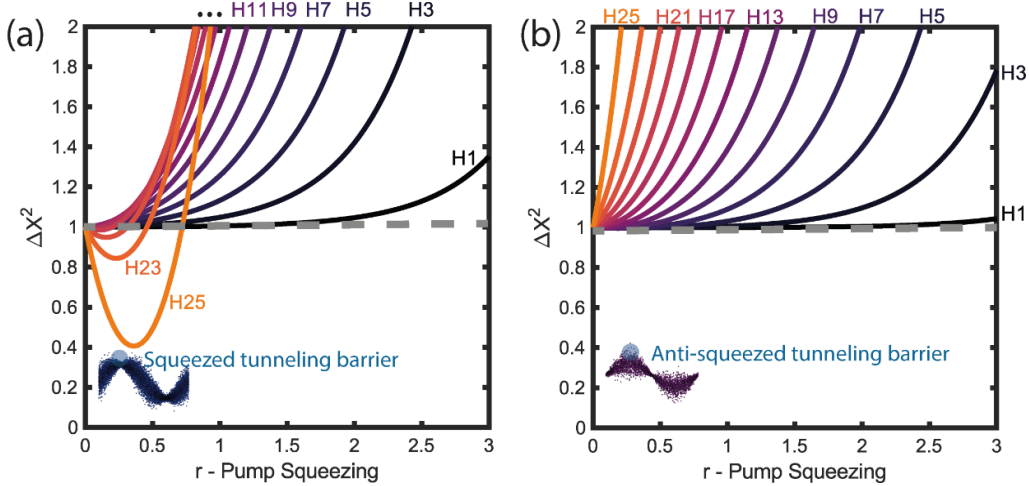

*Figure 2.* **High harmonic quadrature variances for squeezed coherent light.** We evaluate Eq. (11) using the parameters $\theta = 0$ and $4\epsilon^{(1)^4} N^2 q \langle d_{\gamma x}(q\omega)\rangle = 5 \times 10^{-6}\,\text{a.u.}$ (a) Amplitude squeezed pump ($E_{\gamma_x} = 0.0756\,\text{a.u.}, E_{\gamma_y} = 0$). (b) Phase squeezed pump ($E_{\gamma_y} = 0.0756\,\text{a.u.}, E_{\gamma_x} = 0$). For high pump squeezing, none of the harmonics exhibit squeezing because the anti-squeezed quadratures of the pump are mapped to both quadratures of the harmonics. For low pump squeezing, harmonic squeezing occurs when the pump's fluctuations are squeezed during the ionization time.

**High harmonic generation driven by bright squeezed vacuum**

As discussed in the introduction, bright squeezed vacuum (BSV) is the state of light generated by degenerate spontaneous parametric down conversion [52]. In this section, we show that unlike the case of squeezed coherent light, the quantum state of HHG driven by BSV does not resemble the quantum state of the pump whatsoever. Nonetheless, the structure it takes is intuitive, and can be potentially engineered, choosing appropriately material and beam parameters.

We consider a model atom based on a truncated harmonic oscillator potential, for which the HHG emission driven by a coherent state, ($\langle d_\alpha(q\omega)\rangle$ and $\chi^{(q)}(\alpha)$), is known analytically and was derived in Eq. (22) of Ref. [8]. The potential is given by $V\left(|x| < \sqrt{2\beta/\alpha^2}\right) = \frac{\alpha x^2}{2} - \beta$ and $V\left(|x| > \sqrt{2\beta/\alpha^2}\right) = 0$, resulting in an ionization potential $-I_p = -\beta + \frac{3\alpha}{2}$. An exemplary HHG spectrum is plotted in Figure 3(c), with the parameters $I_p = 11\hbar\omega$, $\beta = 50\hbar\omega$, and the ponderomotive energy of the pump is $U_p = 9\hbar\omega$ ($\hbar\omega$ is the laser photon energy). Figure 3(d). shows the scaling of harmonic 23, $|\langle d_\alpha(23\omega)\rangle|^2 \propto |\chi^{(23)}(\alpha)|^2$ with the ponderomotive energy $U_p \propto |E_\alpha|^2 \propto |\alpha|^2$, where $E_\alpha$ and $\alpha$ are the electric field amplitude and coherent state parameter of the driving coherent state field, respectively. In conjunction with Eq. (9), the semi-classical nonlinear response function presented in Figure 3(d) can be used to calculate the quantum state of the harmonics, when driven by BSV. We consider a squeezed vacuum field characterized by the quasi-probability distribution

$$Q_\omega^{(SV)}(\alpha) = \frac{1}{\pi \cosh(r)} \exp\left[-\frac{2\alpha_y^2}{1+e^{2r}} - \frac{2\alpha_x^2}{1+e^{-2r}}\right] \qquad (12)$$

The intensity of the beam is denoted by $I_{vac} = c\hbar\omega \sinh^2(r)/V \equiv \frac{1}{2}\epsilon_0 c|E_{vac}|^2$, where $r$ is the dimensionless squeezing parameter, $V$ is the quantization volume, $\omega = 0.057$ a.u. is the driving frequency (800nm wavelength), and $|E_{vac}| = 0.189$ a.u. is the electric field amplitude of an equally intense coherent state ($1.25 \times 10^{15} \, Watt/cm^2$ intensity). The Husimi distribution $Q(\alpha(\chi))$ is plotted in Figure 3(b), showing an intricate structure, significantly deviating from the Gaussian shape of the pump. Most prominently, it exhibits a sharp, narrow peak centered around $\chi_x = \chi_y = 0$. This is because the majority of coherent state components in $Q_\omega^{(SV)}(\alpha)$ do not generate high harmonics, as their electric field amplitude is too weak (Fig. 3(d)). The rest of the quasi-probability distribution follows a structure similar to a trail winding down a mountain, which peaks at $\chi_{x,y} = 0$. The trail winds because of the circular motion of $\chi$ in the complex plane as $|\alpha|$ is increased. We can also observe that the trail consistently moves down and away from the peak of the mountain. The trail moves away with an exponential rate, since $|\chi|$ increases exponentially with $|\alpha|$ (Fig. 3(d)). The trail moves down because the Husimi distribution $Q_\omega^{(SV)}(\alpha)$ of the squeezed vacuum decreases with $|\alpha|$ (Eq. (12)). Finally, the width of the trail is linear in the $|\alpha_x|$ of $Q_\omega^{(SV)}(\alpha)$.

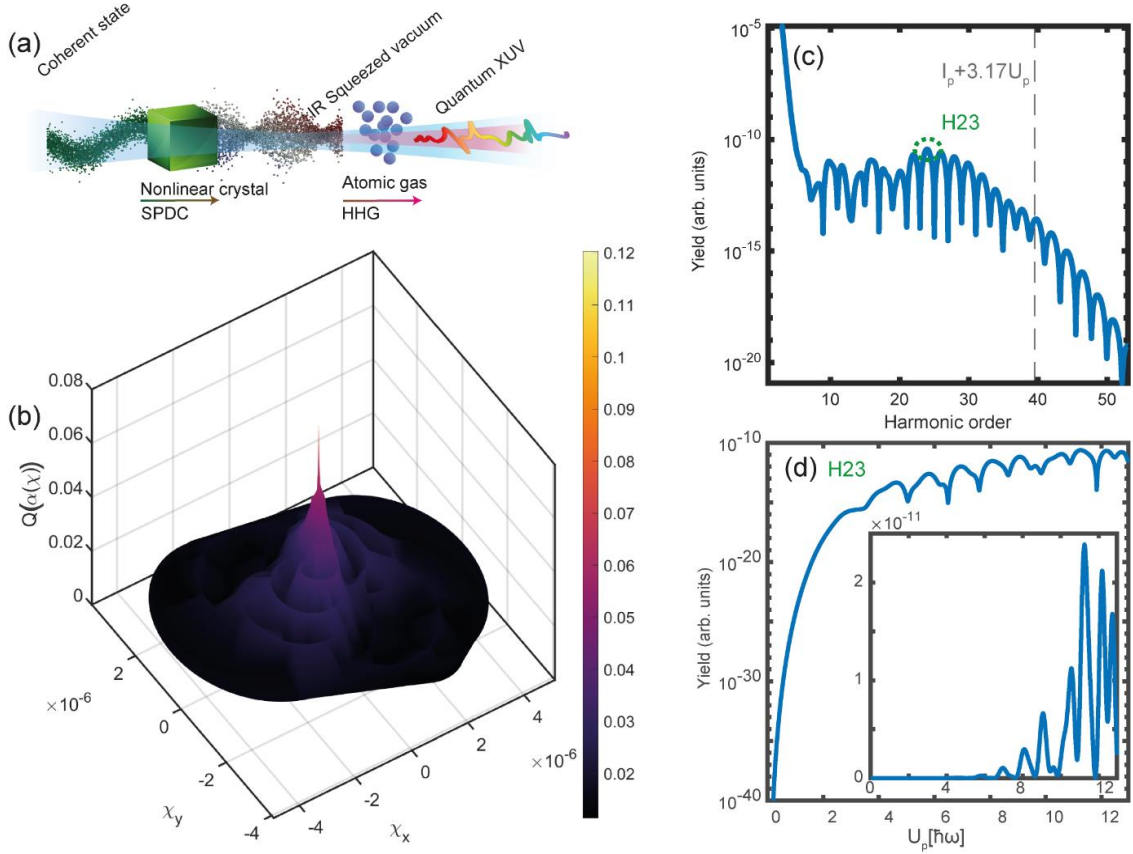

*Figure 3. Quantum state of HHG driven by squeezed vacuum.* **(a)** Illustration of high harmonic generation driven by squeezed vacuum. **(b)** The Husimi distribution $Q(\alpha(\chi))$ of harmonic 23, calculated using $\chi^{(23)}(\alpha)$ (see (d)) and Eq.(12). **(c)** A semi-classical HHG spectrum calculated analytically from a truncated quadratic potential with the ionization potential and ponderomotive energy being $I_p = 11\hbar\omega$ and $U_p = 9\hbar\omega$, *respectively.* **(d)** The scaling of harmonic 23 $\propto \left|\chi^{(23)}(\alpha)\right|^2$ as a function of $U_p \propto |E_\alpha|^2 \propto |\alpha|^2$.

## Conclusion

To summarize, we have explored the quantum state of the emitted light in high harmonic generation in atomic gases, considering the quantum state of the driving field and the semi-classical nonlinear response function of the interacting system. We have shown that the quantum state of the driving field is mapped to the harmonics through the classical nonlinear response function of the system and its Jacobian. We then studied in detail two experimentally feasible cases, high harmonic generation driven by squeezed coherent light and by squeezed vacuum light. For the case of squeezed coherent light, we derived an explicit equation for the quadrature variance of the harmonics, which reveals that the high harmonics inherit squeezing from the pump up to a finite value of pump squeezing. Above this finite value, absolute squeezing of the harmonics does not occur due to the large fluctuations in the anti-squeezed quadrature of the pump. These fluctuations are mapped to fluctuations in both quadratures of the harmonics while the relative quadrature squeezing is always maintained. However, at moderate pump squeezing, transferring squeezing from the pump to the harmonics should be possible. For the case of squeezed vacuum, we have found the quantum state of the emission to be significantly different from the quantum state of the pump. Potentially, our results pave the way towards squeezed high harmonic light, which may bring below-shot-noise metrology to the ultrafast timescale and to the x-ray spectral range. Therefore, we believe our results will have a large impact on HHG, attosecond metrology, quantum optics, and more.

**Bibliography**

[1]  A. McPherson, G. Gibson, H. Jara, U. Johann, T. S. Luk, I. A. McIntyre, K. Boyer, and C. K. Rhodes, *Studies of Multiphoton Production of Vacuum-Ultraviolet Radiation in the Rare Gases*, Journal of the Optical Society of America B **4**, 595 (1987).

[2]  M. Ferray, A. L'Huillier, X. F. Li, L. A. Lompre, G. Mainfray, and C. Manus, *Multiple-Harmonic Conversion of 1064 Nm Radiation in Rare Gases*, Journal of Physics B: Atomic, Molecular and Optical Physics **21**, L31 (1988).

[3]  T. T. Luu, Z. Yin, A. Jain, T. Gaumnitz, Y. Pertot, J. Ma, and H. J. Wörner, *Extreme–Ultraviolet High–Harmonic Generation in Liquids*, Nat Commun **9**, 3723 (2018).

[4]  S. Ghimire, A. D. Dichiara, E. Sistrunk, P. Agostini, L. F. Dimauro, and D. A. Reis, *Observation of High-Order Harmonic Generation in a Bulk Crystal*, Nat Phys **7**, 138 (2011).

[5]  B. Dromey et al., *High Harmonic Generation in the Relativistic Limit*, Nat Phys **2**, 456 (2006).

[6]  P. M. Paul, E. S. Toma, P. Breger, G. Mullot, F. Augé, P. Balcou, H. G. Muller, and P. Agostini, *Observation of a Train of Attosecond Pulses from High Harmonic Generation*, Science (1979) **292**, 1689 (2001).

[7]  P. B. Corkum, *Plasma Perspective on Strong Field Multiphoton Ionization*, Phys Rev Lett **71**, 1994 (1993).

[8]  M. Lewenstein, P. Balcou, M. Y. Ivanov, A. L'Huillier, and P. B. Corkum, *Theory of High-Harmonic Generation by Low-Frequency Laser Fields*, Physical Review A **49**, 2117 (1994).

[9]  J. L. Krause, K. J. Schafer, and K. C. Kulander, *High-Order Harmonic Generation from Atoms and Ions in the High Intensity Regime*, Phys Rev Lett **68**, 3535 (1992).

[10] D. M. Villeneuve, *Attosecond Science*, Contemp Phys **59**, 47 (2018).

[11] N. Dudovich, O. Smirnova, J. Levesque, Y. Mairesse, M. Y. Ivanov, D. M. Villeneuve, and P. B. Corkum, *Measuring and Controlling the Birth of Attosecond XUV Pulses*, Nat Phys **2**, 781 (2006).

[12] D. Shafir, H. Soifer, B. D. Bruner, M. Dagan, Y. Mairesse, S. Patchkovskii, M. Y. Ivanov, O. Smirnova, and N. Dudovich, *Resolving the Time When an Electron Exits a Tunnelling Barrier*, Nature **485**, 343 (2012).

[13] O. Pedatzur et al., *Attosecond Tunnelling Interferometry*, Nat Phys **11**, 815 (2015).

[14] E. Goulielmakis et al., *Real-Time Observation of Valence Electron Motion*, Nature **466**, 739 (2010).

[15] M. Schultze et al., *Attosecond Band-Gap Dynamics in Silicon*, Science (1979) **346**, 1348 (2014).

[16] M. Ossiander et al., *Attosecond Correlation Dynamics*, Nat Phys **13**, 280 (2017).

[17] U. S. Sainadh et al., *Attosecond Angular Streaking and Tunnelling Time in Atomic Hydrogen*, Nature **568**, 75 (2019).

[18] K. T. Kim, C. Zhang, A. D. Shiner, B. E. Schmidt, F. Légaré, D. M. Villeneuve, and P. B. Corkum, *Petahertz Optical Oscilloscope*, Nat Photonics **7**, 958 (2013).

[19] R. E. F. Silva, Jiménez-Galán, B. Amorim, O. Smirnova, and M. Ivanov, *Topological Strong-Field Physics on Sub-Laser-Cycle Timescale*, Nat Photonics **13**, 849 (2019).

[20] O. Neufeld, D. Ayuso, P. Decleva, M. Y. Ivanov, O. Smirnova, and O. Cohen, *Ultrasensitive Chiral Spectroscopy by Dynamical Symmetry Breaking in High Harmonic Generation*, Phys Rev X **9**, 031002 (2019).

[21] D. Ayuso, O. Neufeld, A. F. Ordonez, P. Decleva, G. Lerner, O. Cohen, M. Ivanov, and O. Smirnova, *Synthetic Chiral Light for Efficient Control of Chiral Light–Matter Interaction*, Nat Photonics **13**, 866 (2019).

[22] O. Neufeld and O. Cohen, *Background-Free Measurement of Ring Currents by Symmetry Breaking High Harmonic Spectroscopy*, Phys. Rev. Lett. **123**, 103202 (2019).

[23] M. E. Tzur, O. Neufeld, A. Fleischer, and O. Cohen, *Selection Rules for Breaking Selection Rules*, New J Phys **23**, (2021).

[24] M. E. Tzur, O. Neufeld, E. Bordo, A. Fleischer, and O. Cohen, *Selection Rules in Symmetry-Broken Systems by Symmetries in Synthetic Dimensions*, Nat Commun **13**, 1312 (2022).

[25] P. Kumar, *Quantum Frequency Conversion*, Opt. Lett. **15**, 1476 (1990).

[26] X. Han, W. Fu, C.-L. Zou, L. Jiang, and H. X. Tang, *Microwave-Optical Quantum Frequency Conversion*, Optica **8**, 1050 (2021).

[27] A. Pizzi, A. Gorlach, N. Rivera, A. Nunnenkamp, and I. Kaminer, *Light Emission from Strongly Driven Many-Body Systems*, Nat Phys **19**, 551 (2023).

[28] J. Sloan, A. Gorlach, M. Even Tzur, N. Rivera, O. Cohen, I. Kaminer, and M. Soljačić, *Entangling Extreme Ultraviolet Photons through Strong Field Pair Generation*, ArXiv:2309.16466 (2023).

[29] M. Even Tzur and O. Cohen, *Motion of Charged Particles in Bright Squeezed Vacuum*, To Be Published in Light: Science and Applications (2023).

[30] N. Tsatrafyllis, I. K. Kominis, I. A. Gonoskov, and P. Tzallas, *High-Order Harmonics Measured by the Photon Statistics of the Infrared Driving-Field Exiting the Atomic Medium*, Nat Commun **8**, 15170 (2017).

[31] M. Lewenstein, M. F. Ciappina, E. Pisanty, J. Rivera-Dean, P. Stammer, T. Lamprou, and P. Tzallas, *Generation of Optical Schrödinger Cat States in Intense Laser–Matter Interactions*, Nat Phys **17**, 1104 (2021).

[32] M. Lewenstein et al., *Attosecond Physics and Quantum Information Science*, ArXiv:2208.14769 (2022).

[33] J. Rivera-Dean, P. Stammer, A. S. Maxwell, Th. Lamprou, P. Tzallas, M. Lewenstein, and M. F. Ciappina, *Light-Matter Entanglement after above-Threshold Ionization Processes in Atoms*, Physical Review A **106**, 063705 (2022).

[34] P. Stammer, *Theory of Entanglement and Measurement in High-Order Harmonic Generation*, Phys. Rev. A **106**, L050402 (2022).

[35] P. Stammer, J. Rivera-Dean, A. Maxwell, T. Lamprou, A. Ordóñez, M. F. Ciappina, P. Tzallas, and M. Lewenstein, *Quantum Electrodynamics of Ultra-Intense Laser-Matter Interactions*, ArXiv:2206.04308 (2022).

[36] J. Rivera-Dean, P. Stammer, E. Pisanty, T. Lamprou, P. Tzallas, M. Lewenstein, and M. F. Ciappina, *New Schemes for Creating Large Optical Schrödinger Cat States Using Strong Laser Fields*, J Comput Electron **20**, 2111 (2021).

[37] Philipp Stammer, Javier Rivera-Dean, Theocharis Lamprou, Emilio Pisanty, Marcelo F. Ciappina, Paraskevas Tzallas, and Maciej Lewenstein, *High Photon Number Entangled States and Coherent State Superposition from the Extreme Ultraviolet to the Far Infrared*, Phys. Rev. Lett **128**, 123603 (2022).

[38] J. Rivera-Dean, T. Lamprou, E. Pisanty, P. Stammer, A. F. Ordóñez, A. S. Maxwell, M. F. Ciappina, M. Lewenstein, and P. Tzallas, *Strong Laser Fields and Their Power to Generate Controllable High-Photon-Number Coherent-State Superpositions*, Physical Review A **105**, 033714 (2022).

[39] A. S. Maxwell, L. B. Madsen, and M. Lewenstein, *Entanglement of Orbital Angular Momentum in Non-Sequential Double Ionization*, Nat Commun **13**, 4706 (2022).

[40] A. Gorlach, O. Neufeld, N. Rivera, O. Cohen, and I. Kaminer, *The Quantum-Optical Nature of High Harmonic Generation*, Nat Commun **11**, 4598 (2020).

[41] A. Jechow, M. Seefeldt, H. Kurzke, A. Heuer, and R. Menzel, *Enhanced Two-Photon Excited Fluorescence from Imaging Agents Using True Thermal Light*, Nat Photonics **7**, 973 (2013).

[42] M. Manceau, K. Y. Spasibko, G. Leuchs, R. Filip, and M. V. Chekhova, *Indefinite-Mean Pareto Photon Distribution from Amplified Quantum Noise*, Phys Rev Lett **123**, 123606 (2019).

[43]  Y. Qu and S. Singh, *Photon Correlation Effects in Second Harmonic Generation*, Opt Commun **90**, 111 (1992).

[44]  I. N. Agafonov, M. V Chekhova, and G. Leuchs, *Two-Color Bright Squeezed Vacuum*, Phys Rev A (Coll Park) **82**, 11801 (2010).

[45]  T. S. Iskhakov, A. M. Pérez, K. Y. Spasibko, M. V Chekhova, and G. Leuchs, *Superbunched Bright Squeezed Vacuum State*, Opt Lett **37**, 1919 (2012).

[46]  A. M. Pérez, T. S. Iskhakov, P. Sharapova, S. Lemieux, O. V Tikhonova, M. V Chekhova, and G. Leuchs, *Bright Squeezed-Vacuum Source with 1.1 Spatial Mode*, Opt Lett **39**, 2403 (2014).

[47]  M. A. Finger, T. S. Iskhakov, N. Y. Joly, M. V Chekhova, and P. S. J. Russell, *Raman-Free, Noble-Gas-Filled Photonic-Crystal Fiber Source for Ultrafast, Very Bright Twin-Beam Squeezed Vacuum*, Phys Rev Lett **115**, 143602 (2015).

[48]  P. R. Sharapova, G. Frascella, M. Riabinin, A. M. Pérez, O. V Tikhonova, S. Lemieux, R. W. Boyd, G. Leuchs, and M. V Chekhova, *Properties of Bright Squeezed Vacuum at Increasing Brightness*, Phys Rev Res **2**, 13371 (2020).

[49]  A. Gorlach, M. E. Tzur, M. Birk, M. Krüger, N. Rivera, O. Cohen, and I. Kaminer, *High-Harmonic Generation Driven by Quantum Light*, Nat Phys (2023).

[50]  M. Even Tzur, M. Birk, A. Gorlach, M. Krüger, I. Kaminer, and O. Cohen, *Photon-Statistics Force in Ultrafast Electron Dynamics*, Nat Photonics **17**, 501 (2023).

[51]  O. H. Heckl et al., *High Harmonic Generation in a Gas-Filled Hollow-Core Photonic Crystal Fiber*, Applied Physics B **97**, 369 (2009).

[52] T. S. Iskhakov, A. M. Pérez, K. Y. Spasibko, M. V Chekhova, and G. Leuchs, *Superbunched Bright Squeezed Vacuum State*, Opt Lett **37**, 1919 (2012).

[53] M. S. Kim, F. A. M. de Oliveira, and P. L. Knight, *Properties of Squeezed Number States and Squeezed Thermal States*, Physical Review A **40**, 2494 (1989).

[54] J. Rivera-Dean, T. Lamprou, E. Pisanty, P. Stammer, A. F. Ordóñez, A. S. Maxwell, M. F. Ciappina, M. Lewenstein, and P. Tzallas, *Strong Laser Fields and Their Power to Generate Controllable High-Photon-Number Coherent-State Superpositions*, Physical Review A **105**, (2022).

[55] M. D. Feit, J. A. Fleck Jr., and A. Steiger, *Solution of the Schrödinger Equation by a Spectral Method*, J Comput Phys **47**, 412 (1982).

[56] M. G. A. Paris, *Displacement Operator by Beam Splitter*, Phys Lett A **217**, 78 (1996).

[57] C. Riek, D. v. Seletskiy, A. S. Moskalenko, J. F. Schmidt, P. Krauspe, S. Eckart, S. Eggert, G. Burkard, and A. Leitenstorfer, *Direct Sampling of Electric-Field Vacuum Fluctuations*, Science (1979) **350**, 420 (2015).